\def \figurecaption            {\noindent\hangindent0.5in\hangafter=1}
\def \um             {\ifmmode \mu m\else $\mu$m\fi}
\def \HOH            {\ifmmode H_2O\else H$_2$O\fi}
\def \HH             {\ifmmode H_2\else H$_2$\fi}
\def \ch             {\ifmmode CH_4\else CH$_4$\fi}
\def \nuthree        {$\nu_3$}

\documentstyle[aasms4,epsf]{article}
\begin{document}

\title  {The Onset of Methane in L Dwarfs}

\author{ Keith S.~Noll}
\affil{Space Telescope Science Institute, 3700 San Martin Dr.,
Baltimore, MD 21218; noll@stsci.edu}

\author{ T.~R.~Geballe}
\affil{Gemini North}

\author{ S.~K.~Leggett}
\affil{Joint Astronomy Centre}

\author{ Mark S.~Marley}
\affil{New Mexico State University}

\begin{abstract}  

We have detected weak absorption features produced by the strong
$\nu_3$ methane band at 3.3~\um\ in two L dwarfs,
2MASSW~J1507476-162738 and 2MASSI~J0825196+211552, classified by
Kirkpatrick et al. (2000) as spectral types L5 and L7.5 respectively. 
These absorptions occur in objects warmer than any in which methane
previously has been detected, and mark the first appearance of methane
in the ultracool star -- to -- brown dwarf spectral sequence.

\end{abstract}


\section{Introduction}

The growing population of ultracool stars and substellar objects has been
divided observationally into two new spectral classifications
(Kirkpatrick et al. 1999): T dwarfs 
(sometimes called methane dwarfs) which have methane absorption in the near
infrared (1-2.5 \um) (e.g., Oppenheimer et al. 1995; Geballe et al. 1996;
Strauss et al. 1999, Burgasser et al. 1999, Tsvetanov et al. 2000)  and
the warmer L dwarfs with no detectable near-IR methane (e.g., Kirkpatrick et
al.~1999, Martin et al.~1999). In the transition from L to T the
abundance of carbon monoxide decreases as methane becomes more stable
against collisional dissociation, the overwhelming abundance of hydrogen
drives chemical equilibrium toward higher abundances of methane, and the
vibration-rotation bands of \ch , along with those of H$_{2}$O, dominate
the infrared spectrum of the object.  However, the details of the
transition, and in particular its rapidity as the effective temperature
decreases have been unclear. 

Until recently no objects bridging the distinct CO/\ch\ spectral
separation of the L and T dwarfs had been found.
An estimate by Kirpatrick et al.~(2000) of the
effective temperature of the L8 dwarf Gl~584C (2MASSW~J1523226+301456),
based on an assumed bolometric correction and comparison to Gl~229B,
gave $T_{eff}~\sim$~1300~K, which led those authors to speculate that
the gap in temperature between L dwarfs and known T dwarfs ``must
be {\em much} smaller than 350~K and possibly smaller than 100~K''. 
The sudden appearance of strong near-infrared methane over such a small
decrement in temperature would imply special atmospheric conditions
such as clouds suddenly forming or clearing and would explain the lack
of transitional objects.  However, we note that Basri et al.~(2000)
find significantly higher temperatures, $T_{eff} \sim$ 1700~K, for the
coolest L dwarfs. This discrepancy demonstrates that the transition
from L to T dwarfs remains the least well constrained portion of the
brown dwarf sequence.

The recent detection of three dwarfs with weak but detectable methane
bands in the 1.0-2.5~$\mu$m region (Leggett et al.~2000) now has begun
to populate the gap between L dwarfs and the previously known
methane-rich T dwarfs, and suggests that observational selection
effects and/or small number statistics may instead be the explanation
of the missing transition objects. Leggett et al. and Kirkpatrick et
al.~(1999) both propose that the T spectral sequence begin with the
appearance of \ch\ at 1.0-2.5~\um .  Regardless of the details of
classification, the onset of methane absorption at any wavelength and
the range of temperatures and spectral subclasses over which \ch\
increases in abundance and becomes dominant over CO are important
diagnostics for the physical characterization of these objects.

In order to search for the earliest appearance of methane in the
spectrum of cool dwarfs, we have conducted observations of L dwarfs
near 3.3 \um\ where the strong \nuthree\ band of \ch\ should first
appear.  This band has not been seen previously in any L dwarf. It is
very deep and broad in the T dwarf Gl~229B (Oppenheimer et al. 1998)
and should be strong in the transition objects observed by Leggett et
al. (2000) and other T dwarfs, but should disappear somewhere within
the L spectral sequence.  Our observations are part of an ongoing
program to obtain spectra of L and late-M dwarfs using the United
Kingdom Infrared Telescope's (UKIRT) Cooled Grating Spectrometer 4
(CGS4, Mountain et al. 1990).  In this {\sl Letter} we report the
detection of weak absorption from the \nuthree\ band of \ch\ in two L
dwarfs, 2MASSW~J1507476-162738 and 2MASSI~J0825196+211552.

\section{Observations and Analysis}

On UT 2000 May 22 and 23 we used UKIRT and CGS4 with its 40 l/mm grating,
2-pixel (1.2 arcsec) wide slit, and 300 mm focal length camera to obtain
3.15-3.78~$\mu$m spectra of several L dwarfs. Pertinent details of the
observations are given in Table 1. The above instrumental configuration
yields a resolving power of $R\approx~600$. The array was read out once,
at the end of each exposure (which was typically 0.5-1.5 seconds duration,
depending on the target brightness), and then reset; multiple exposures
were combined in a preprocesser to obtain a single spectral frame. After
the first integration the array was translated in the dispersion direction
by one pixel and the multiple exposures repeated; this ensured that all
wavelengths were observed by at least one good pixel. Following the two
integrations the telescope was nodded along the slit by 7.2 arcsec and the
observing procedure repeated, in order to facilitate sky subtraction while
maximizing the time on our source.

An average of pairs of differenced, nodded spectral images was produced by
the CGS4 on-line data reduction program. The individual images making up
this average spectral image were flatfielded and bad pixels masked off by
the on-line program.  Spectra were reduced further off-line using the
FIGARO software package and following standard steps for CGS4 data.  
These steps include extraction of source spectra from designated array
rows, wavelength calibration using argon lamp lines, and removal of
telluric lines and flux calibration using similarly prepared spectra of
standard stars.

The fully reduced spectra of three L dwarfs, 2MASSW~1506544+132106, and
2MASSW~J1507476-162738, and 2MASSI~J0825196+211552, (hereafter 2M1506,
2M1507, and 2M0825) are shown in Figure 1. Based on the strength of
features in the 0.8-1.0 \um\ portion of the spectrum, they have been
classified L3 (2M1506), L5 (2M1507), and L7.5 (2M0825) by Kirkpatrick et al.
(2000). The spectrum of 2M0825 has been rebinned by a factor of four to
reduce the noise at the expense of lower resolution.  Each spectrum has
been normalized to its median flux and offset to allow for 
comparison of the relative strengths of absorption features.

No sign of a spectral absorption near 3.3~\um\ is seen in 2M1506, but a
weak feature of width $\sim$0.03~$\mu$m is present in 2M1507, and a
stronger, although noisier, feature is present in 2M0825.  Both the
positions and shapes of the spectral features match that expected for
the Q branch of the \nuthree\ band of \ch . The wavelength of the
observed feature extends longward from the strong telluric Q branch
absorption which occurs near 3.316~$\mu$m. The low lying rotational
levels of the ground state of methane are heavily populated in the
earth's atmosphere, rendering the narrow interval 3.312-3.323~$\mu$m
virtually unobservable (see Fig.~2). In warmer objects,
however, higher J levels are populated and the Q branch becomes
significantly broader than the terrestrial band, extending mainly to
longer wavelengths, and is detectable from the ground.  The
correspondence of a stronger absorption feature with a cooler spectral
classification is in accord with expectations for \ch , which is
chemically favored at lower temperatures (Fegley \& Lodders 1996). 
Finally, the shape of the observed spectral feature agrees
qualitatively with models of methane and water spectra at L dwarf
temperatures as discussed below.   Thus, we conclude that the
absorption feature longward of 3.3~\um\ is the \ch\ \nuthree\ Q branch.

In Figure 2 we compare the spectrum of 2M1507 to model atmosphere spectra,
one containing only water vapor and one with water vapor and methane.  
The model was computed using a line-by-line radiative transfer code, an
atmospheric P-T profile for an object with T$_{eff}$=1800 K and g=1000
m/s$^2$, hot-water and methane line lists (Freedman, private
communication), and a linear slope as needed to match to the slope of the
observed spectrum.  The models have been smoothed, normalized, and offset
for comparison.  Given the disagreements in the overall slope, the model
spectra must be regarded as qualitative only, although the identification
of individual spectral features is robust. The weak features in the
spectrum 2M1507 match the spectrum of water except near 3.3 \um\
where additional absorption from \ch\ is present.

\section{Discussion}

In the brown dwarf literature much of the discussion about the
transition from CO to \ch\ has focussed on the temperature at which
atmospheric abundances of CO and \ch\ in thermochemical equilibrium are
equal.   Authors have tended to equate this temperature with the
effective temperature of the brown dwarf in making predictions of the
location in the spectral sequence where methane will become detectable
in the spectrum.  While a useful shorthand, this approach greatly
oversimplifies the question of \ch\ observability in brown dwarfs. 
First, it ignores the large differences in strengths between methane
bands; the more frequently observed near-IR bands are two orders of
magnitude weaker than the fundamental at 3.3 \um\ that is the focus of
this work.  Second, even in dwarfs with effective temperatures above
the \ch -CO equilibrium temperature ($\sim$1400~K at P=10 bar,
$\sim$1100~K at P=1bar), there exist overlying cooler layers where \ch\
is more abundant.  Third, the dropoff of \ch\ abundance below the
equilibrium point is slow so that even at relatively high temperatures
the \ch\ abundance remains above 10$^{-5}$ (Fegley \& Lodders 1996;
Burrows \& Sharp 1999).

It is important to emphasize that \ch\ will be observable in the
spectrum of an object when there is a sufficiently large column
abundance above the optically thick lower boundary of the atmosphere as
determined by cloud, \HH\ continuum, and/or line opacities.  This
condition can be met for a wide range of possible model atmospheres and
chemical profiles. Because of the strength of the $\nu_3$ fundamental
band, emission within the band originates very high in the atmosphere.
For a solar abundance of carbon entirely in methane, the band is
already optically thick by 1 mbar in a 30 Jupiter mass object.   The
temperatures at such pressures are always well below the effective
temperature of the brown dwarf (Marley 2000).  For a gray atmosphere
this region of the atmosphere would be near $0.84T_{eff}$.  Brown
dwarf atmospheres, however, are far from gray and the upper atmosphere
is much cooler.  For the cloud-free model used to generate Figure 2,
for example, the temperature at 1 mbar is closer to $0.6T_{eff}$ or
$1080\,\rm K$. While still not within the methane stability region at
this pressure (Fegley \& Lodders 1996) the equilibrium methane
abundance under such conditions is not negligible and is larger than
that found in air at the effective temperature, $T_{eff}$, of the
brown dwarf. Dusty atmospheres are warmer at a given pressure, but the
air temperature at $P< 1\,\rm bar$ is still well below the effective
temperature.

Although the models presented in Fig.~2 are qualitative in the sense
that no attempt is made to match the absolute flux levels or slope of
the spectrum, models covering a range of effective temperatures give
some insight into factors that control the appearance of \ch\ (Noll et
al. 2000).  The effective temperature has little direct influence on
the appearance of the model spectrum of either water or methane in the
range 1400-2000~K; the relative strengths of features change slowly
over that range. Thus, the choice of an 1800~K model P-T profile in
Fig.~2 is not necessarily indicative of the effective temperature of
2M1507.  The appearance of the spectrum at 3~$\mu$m is dictated much
more strongly by the methane abundance profile as a function of
temperature and pressure.  Our models used the thermochemical abundance
of methane predicted for a cloud-free atmosphere model by the chemistry
of Burrows \& Sharp (1999).   In the model shown we have reduced the
predicted methane abundance at each level by one-third (corresponding to
somewhat higher temperatures) to better match
the depth of the observed band.  More realistic dusty atmosphere models
are indeed warmer and have less methane at a given pressure level, but
such models introduce many more variables than we wish to consider
here.  

Considerable confusion exists in characterizing the portion of the
spectral sequence corresponding to the onset of methane. For example,
Basri et al.~(2000) find an effective temperature of $T_{eff}$~=~1750~K
for DENIS-P J0205-1159 from an analysis of CsI and RbI lines in its
spectrum and, based on this temperature, classify it as an L5 dwarf. 
For DENIS-P J1228-1547 they obtain $T_{eff}$ = 1800~K, leading to a
classification as an L4.5 dwarf.   However, based on spectral
morphology,  Kirkpatrick et al.~1999 classify DENIS-P J0205-1159 as L7
and advocate an effective temperature closer to 1400~K.  Similar
discrepancies exist for other late L dwarfs as summarized by Martin et
al.~(1999) who offer a conversion from their classification scheme to
that of Kirkpatrick et al.~(1999,2000).  

Despite the caveats noted above, the weakness of the methane band in
the L5 (by the Kirkpatrick scheme) object 2M1507 tends to support a
temperature considerably higher than 1400~K where all of our models
predict a very strong \nuthree\ methane band.  This same statement
applies to the L7.5 object 2M0825 where the weakness of the \nuthree\
methane band is all the more surprising.  Kirkpatrick et al. (2000)
would argue for a temperature near 1400 K for this object.  If the
transformation offered by Martin et al. (1999) holds for this object it
would be classified by them as roughly an L5.5 which in the Basri et
al. (2000) scheme corresponds to an effective temperature of $\approx$
1700-1750~K.  If the relative classification of these two objects being
separated by 2 to 2.5 classes is correct, it implies that the onset of
methane in the L dwarfs is a gradual process relative to either the
Kirkpatrick et al. or Martin et al.
classification schemes.  Whether this gradual onset is due to the
spectral classes being separated by only small temperature increments
or whether it is due to the fundamental chemistry of methane in these
objects is unclear.   Further modelling and additional spectra
will help to clarify this situation.

Finally, we comment on the ongoing development of classification schemes
for substellar objects.  Kirkpatrick et al.~(1999) proposed the adoption
of two new spectral classes, L and T, to cover stars and brown dwarfs from
the end of the M dwarf sequence down to objects similar to Gl~229B.  The T
dwarfs, or methane dwarfs as they are sometimes called, are characterized by
the detectability of methane in the near-IR overtone bands in the
1.0-2.5~\um\ region.  As an observational definition, this is as
satisfactory as any other arbitrary definition.  However, the term
``methane dwarf'' should be used cautiously and only within this narrow
definition.  It would be incorrect, as shown by our observations, to make
the inference that the warmer L dwarfs are methane-free.

\section{Conclusions}

We have detected weak absorption due to the \nuthree\ band of methane
in two L dwarfs, 2M1507 and 2M0825, with spectral types L5 and L7.5
respectively (as classified by Kirkpatrick et al.~2000).  These are the
warmest objects in which \ch\ has been observed.  We do not detect \ch\
in the spectrum of the slightly warmer object, 2M1506, classified as
L3.  We conclude that the onset of methane absorption occurs near
spectral type L5 in the spectral sequence.  Preliminary models indicate
that T$_{eff} \approx$1800~K or higher may be required to match the
observed weakness of the 3.3~\um\ fundamental band in both the L5 and
L7.5 dwarfs.  Our results, while preliminary, tend to support effective
temperatures similar to those found by Basri et al.~(2000) for late L
dwarfs, and are higher than those advocated by the classification scheme
of Kirkpatrick et al. (1999), although the role of dust has yet to be
fully explored.  The term ``methane dwarf'', a common alternative to
the proposed designation of T dwarf, should be used with the
understanding that it refers to specific overtone and combination bands
in the near-infrared and not to the composition of the brown dwarf
atmosphere. Likewise, the designation L dwarf does not imply the
undetectability of the strongest \ch\ bands.

\bigskip\bigskip

\noindent{\bf Acknowledgements}

We thank R.~Freedman for help with access to high temperture lines
lists for water and methane. We also wish to thank the staff of the
United Kingdom Infrared Telescope, which is operated by the Joint
Astronomy Centre on behalf of the U.K. Particle Physics and research
Council. This work was supported by NASA grant NAG5-8314 to KSN at the
Space Telescope Science Institute which is operated by the Association
of Universities for Research in Astronomy under NASA contract
NAS5-26555, by a National Research Council Senior Resident Reseach
Associateship at the Marshall Space Flight Center (KSN), and by NSF
CAREER grant AST-9624878 to MSM.

\newpage 

\noindent{\bf References} 

\noindent{Basri, G., Mohanty, S., Allard, F., Hauschildt, P.~H.,
Delfosse, X., Marti\'n, E., Forveille, T., \& Goldman, B. 2000
astro-ph/0003033} 

\noindent{Burgasser, A.~J., et al. 1999 ApJ, 522, L65}

\noindent{Burrows, A. \& Sharp, C.~M. 1999 ApJ, 512, 843}

\noindent{Fegley, M.~B. \& Lodders, K. 1996, ApJ, 472, L37}

\noindent{Geballe, T.~R., Kulkarni, S.~R.,Woodward, C.~E., \& Sloan,
G.~C. 1996, ApJ, 467, L101}

\noindent{Kirkpatrick, J.~D., et al. 1999, ApJ, 519, 802}

\noindent{Kirkpatrick, J.~D., et al. 2000, astro-ph/0003317}

\noindent{Leggett, S.~K., et al. 2000, ApJ, 536, L35}

\noindent{Marley, M. (2000) in From Giant Planets to Cool Stars
(C. Griffith \& M. Marley, eds). ASP Conf. Series, in press}

\noindent{Martin, E.~L., Delfosse, X., Basri, G., Goldman, B.,
Forveille, Th., \& Zapatero Osorio, M. 1999, ApJ, 118, 2466}

\noindent{Mountain, C.~M., Robertson, D., Lee, T.~J. \& Wade, R. 1990,
Proc. SPIE, 1235, 25}

\noindent{Noll, K.~S., Geballe, T.~R., Leggett, S.~K., Marley, M. 2000,
IAU General Assembly XXIV, Ultracool Dwarf Stars, abstract, in press}

\noindent{Oppenheimer, B.~R., Kulkarni, S.~R., Matthews, K. \& Nakajima,
T. 1995, Science, 270, 1478}

\noindent{Oppenheimer, B.~R., Kulkarni, S.~R., Matthews, K. \& van Kerkwijk,
M.~H. 1998, ApJ 502, 932}

\noindent{Tsvetanov, Z.~I., et al. 2000, ApJ, 531, L61}

\newpage

\begin{deluxetable}{ccccc}
\tablecaption{Observing Log \label{tbl-1}}
\tablewidth{0pt}
\tablehead{
\colhead{Object Name} & \colhead{Sp. Type \tablenotemark{1}} &
\colhead{K$_S$ \tablenotemark{1}} & \colhead{Integ. Time}
& \colhead{Calib. Star} \\
 & & \colhead{mag.} & \colhead{min.} & \\
}
\startdata
2MASSI J0825196+211552 & L7.5 & 13.05 &  64 & HR 3625 (F9V) \\
2MASSW J1506544+132106 & L3   & 11.75 & 128 & HR 6181 (F5IV) \\
2MASSW J1507476-162738 & L5   & 11.30 & 120 & HR 5927 (F7V) \\
\tablenotetext{1} {Kirkpatrick et al. 2000}
\enddata
\end{deluxetable}

\newpage

\begin{figure}
\epsfxsize=7.0in
\epsffile{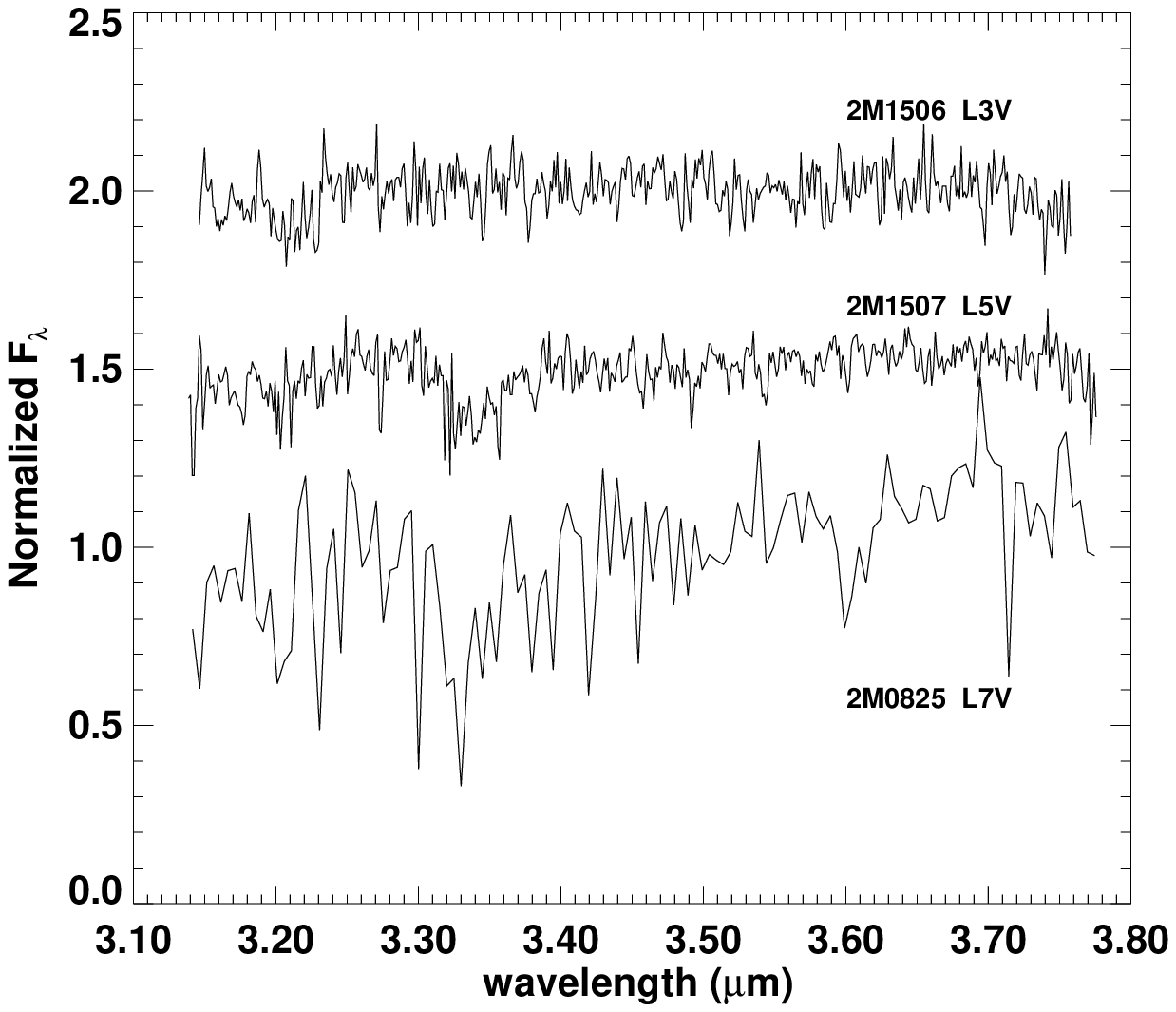}

\figurecaption {{\bf Figure 1.} Spectra of three brown dwarfs, 2M0825,
2M1506, and 2M1507.  The spectra have been rebinned, normalized, and
offset as described in the text.  The broad absorption feature at 3.33
\um\ is the \nuthree\ Q branch of \ch . It is present in both the L5 and
L7.5 objects, but is undetected in the L3.  }

\end{figure}

\newpage

\begin{figure}
\epsfxsize=7.0in
\epsffile{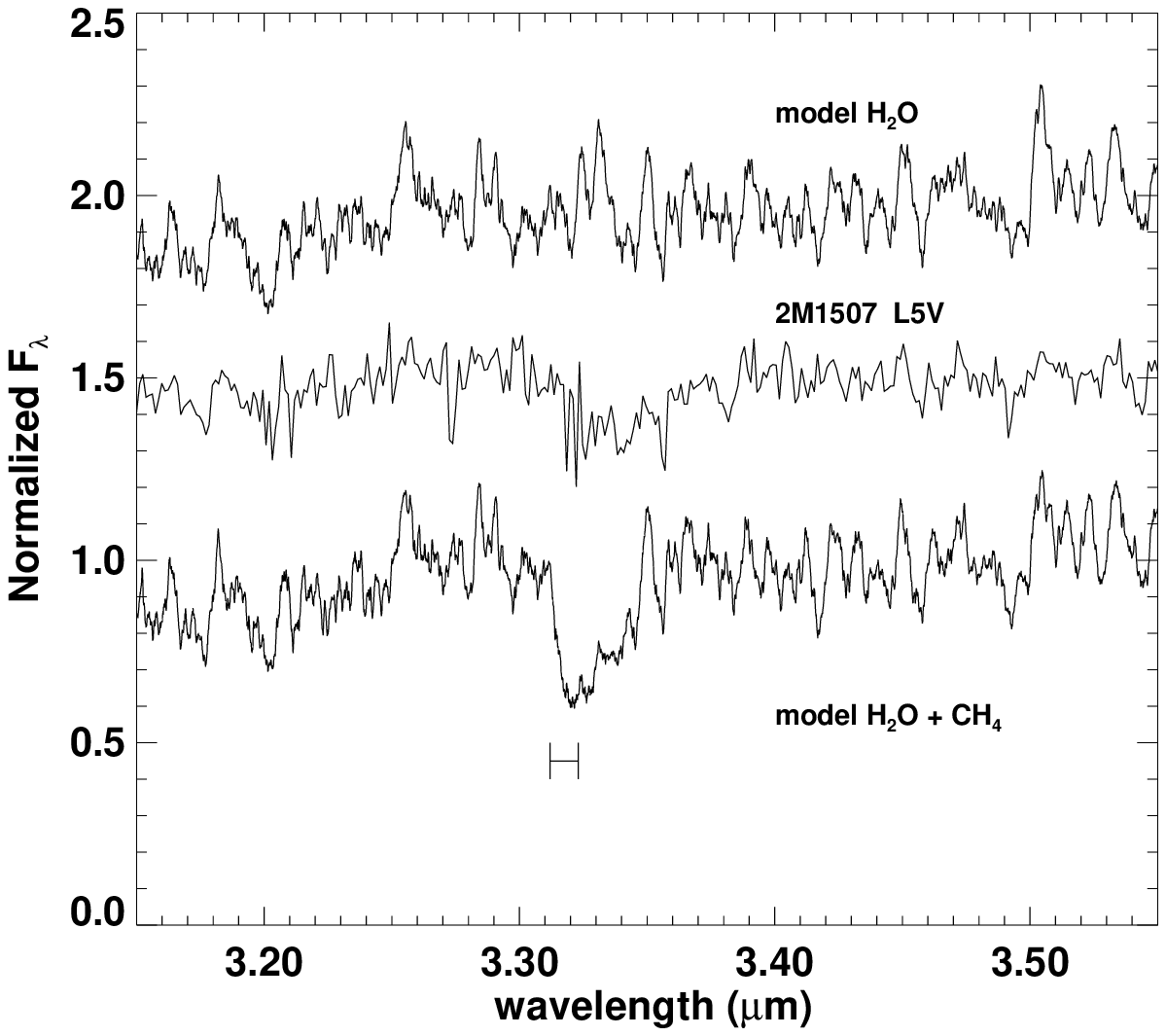}

\figurecaption {{\bf Figure 2.} Two model spectra for a brown dwarf of
effective temperature 1800~K, one with only water vapor and one with
water and methane (at 1/3 the predicted abundance for \ch ), are
compared to the spectrum of 2M1507.  The model spectra have been
smoothed with a 4 cm$^{-1}$ boxcar and normalized to their median
values.  The water model has been offset
by +1.0 and the normalized spectrum of 2M1507 has been offset by +0.5. 
The absorption feature at 3.32-3.35~\um\ in 2M1507 corresponds with the
\ch\ \nuthree\ Q branch. The two vertical lines indicate the interval
where terrestrial \ch\ absorption is strong.}

\end{figure}

\end{document}